\documentclass[12pt]{article}
\usepackage{amsfonts}
\date{}
\title{Isoperiodic deformations of the acoustic operator and periodic solutions
of the Harry Dym equation}
\author{D.V.Zakharov\footnote{Columbia University, New York, USA, {\it
zakharov@math.columbia.edu}}}
\newtheorem{theorem}{Theorem}
\newtheorem{lemma}{Lemma}
\newtheorem{proposition}{Proposition}

\begin{document}
\maketitle

\begin{abstract}
\noindent We consider the problem of describing the possible spectra
of an acoustic operator with a periodic finite-gap density. We construct
flows on the moduli space of algebraic Riemann surfaces that preserve the
periods of the corresponding operator. By a suitable extension of the
phase space, these equations can be written with quadratic
irrationalities.

\end{abstract}

\section{Introduction}
\renewcommand{\theequation}{\thesection.\arabic{equation}}

\setcounter{equation}{0}

In the study of periodic potentials of the one-dimensional Schr\"odinger
operator, we encounter the following problem. A given periodic
potential $u(x)$ of the Schr\"odinger operator
\begin{equation}
L_S=-\frac{d^2}{d y^2} + u(y)
\end{equation}
determines a Riemann surface, called the spectral curve, which is
algebraic if the potential has a finite number of energy gaps. Conversely,
given a hyperelliptic Riemann surface of genus $g$, there exists a potential $u(x)$ with the
corresponding spectrum, which can be extended to a solution $u(x,\vec{t})$
of the KdV hierarchy. However, this potential is not in general
periodic. We cannot extend the direct spectral transform to all quasi-periodic
potentials. For instance, the spectrum of a quasi-periodic potential may
have the structure of a Cantor set (see \cite{Sinai}), making it impossible to
associate a Riemann surface to it. The problem of determining which
Riemann surfaces correspond to periodic finite-gap potentials did not at
first receive much attention.

The periods of a potential of the Schr\"odinger operator are expressed as
certain Abelian integrals on the corresponding spectral curve. The
potential has a given period if and only if these integrals are
integer multiples of the period. Therefore, the subset of curves
corresponding potentials with a given period forms a transcendental
submanifold in the moduli space of hyperelliptic Riemann surfaces.

A description of the possible spectra of a periodic Scr\"odinger operator,
not necessarily finite-gap, was given by Marchenko and Ostrovsky in
\cite{Marchenko} in terms of the properties of a conformal map of a
certain type. However, their approach was difficult from a computational
viewpoint. An effective solution was given by Grinevich and Schmidt in
\cite{Grinevich}, called the method of isoperiodic deformations. Using an
idea developed in \cite{Ercolani}, \cite{Krichever3}, the solution consists of finding a
set of differential equations on the spectral data that do not change the
periods of the solution. The submanifold corresponding to solutions of a
given period is then preserved by these flows. By a suitable extension of
the phase space, the equations of isoperiodic deformation can be written
with a rational right-hand side, making them convenient for numerical
simulation.

This paper extends this approach to the study of the periodic
finite-gap densities acoustic operator
\begin{equation}
L_A=-r^2(x)\frac{d^2}{d x^2}.
\end{equation}
Finite-gap densities of the acoustic operator and the corresponding
periodic solutions of the Harry Dym equation
\begin{equation}
r_{t}=r^3r_{xxx}
\end{equation}
were first constructed by Dmitrieva in \cite{Dmitrieva1},
\cite{Dmitrieva2}, \cite{Dmitrieva3}, using a Hopf transformation to the
Schr\"odinger operator introduced in \cite{Nucci}. Periodic solutions of the Harry Dym equation
were recently shown to be relevant to the Saffman-Taylor problem
(see \cite{Tanveer}), while finite-gap periodic densities of the acoustic
operator are related to geodesics on the ellipsoid (see \cite{Moser}, \cite{Veselov1},
\cite{Veselov2}).

In sections 2 and 3, we recall the spectral theory of the periodic
Schr\"odinger operator and the method of isoperiodic deformations. In
section 4, we derive a spectral theory for the acoustic operator by
relating it to the Schr\"odinger operator, and in section 5 we construct
the equations of isoperiodic deformation for the acoustic operator. By
extending the phase space, we write these equations with quadratic
irrationalities. The principal result of this paper is Theorem 2, which gives
the explicit form of the deformation equations.

\section{Periodic finite-gap potentials of the Schr\"odinger operator}

\setcounter{equation}{0}

We first recall the spectral theory of the one-dimensional periodic finite-gap Schr\"odinger
operator
\begin{equation}
L_S=-\frac{d^2}{d y^2} + u(y),
\end{equation}
where $u(y)$ is a smooth periodic real-valued potential with period $\Pi$.
For references, see \cite{FourAuthors}.

We consider two spectral problems for $L_S$:
\begin{enumerate}
\item The standard problem in $L^2(\mathbb{R})$:
$$
L_S \varphi=\lambda\varphi,\hspace{2mm} |\varphi(y)|<\infty
\hspace{2mm}\mathrm{as}\hspace{2mm}y\rightarrow \pm\infty,
$$

\item The Dirichlet problem on a period:
$$L_S \varphi=\lambda \varphi, \hspace{2mm}
\varphi(y_0)=\varphi(y_0+\Pi)=0.
$$
\end{enumerate}

The spectrum of the first problem is continuous and consists of an infinite number
of segments $[\Lambda_0,\Lambda_1],[\Lambda_2,\Lambda_3],\ldots$, with
$\Lambda_{2j}<\Lambda_{2j+1}\leq\Lambda_{2j+2}$.
The spaces between these segments $(\Lambda_{2j+1},\Lambda_{2j+2})$, which
may be of zero length, are called the energy
gaps. The second problem has a purely discrete spectrum
$d_j(y_0)$, with exactly one eigenvalue inside or on the
boundary of each of the
energy gaps, $d_j(y_0)\in[\Lambda_{2j-1},\Lambda_{2j}]$, including the
degenerate gaps.

The principal case of interest is when the potential $u(y)$ has only a
finite number of energy gaps of non-zero length, such a potential is called
finite-gap. Let
$(-\infty,\lambda_0),(\lambda_1,\lambda_2),\ldots,(\lambda_{2g-1},\lambda_{2g})$ denote
the non-trivial energy gaps, and let $\gamma_j(y_0)$ denote the
eigenvalue of the Dirichlet problem lying in the $j$-th non-trivial gap.

The direct spectral transform assigns to the finite-gap potential $u(y)$
the following data
\begin{enumerate}
\item A hyperelliptic Riemann surface $\Gamma$ of genus $g$ together with a
two-sheeted covering $\lambda:\Gamma\rightarrow\mathbb{C P}^1$ ramified at
$\lambda_0,\ldots,\lambda_{2g}$ and $\infty$

\item A meromorphic function
$\varphi(y,y_0,P)$ on $\Gamma\backslash\lambda^{-1}(\infty)$, called the
Bloch-Floquet function, that has $g$ simple
poles $P_1(y_0),\ldots,P_g(y_0)$ on $\Gamma$ satisfying $\lambda(P_k(y_0))=\gamma_k(y_0)$.
\end{enumerate}
This function is a joint eigenfunction of the Schr\"odinger operator and
the monodromy operator:
\begin{equation}
L_S\varphi(y,y_0,P)=\lambda(P)\varphi(y,y_0,P),
\label{BlochFloquet1}
\end{equation}
\begin{equation}
\varphi(y+\Pi,y_0,P)=\mu(P)\varphi(y,y_0,P),
\label{BlochFloquet2}
\end{equation}
and has the following high-energy expansion:
\begin{equation}
\varphi(y,y_0,P)=\exp(i(y-y_0)\sqrt{\lambda})(1+o(1)).
\label{BlochFloquet3}
\end{equation}
The logarithmic derivative of the Bloch-Floquet function is equal to
\begin{equation}
\chi(y,P)=-i\frac{\varphi_y(y,y_0,P)}{\varphi(y,y_0,P)}
=\frac{\sqrt{R(\lambda(P))}}{S(y,\lambda(P))}
-\frac{i}{2}\frac{S_y(y,\lambda(P))}{S(y,\lambda(P))},
\label{LogBF}
\end{equation}
where the functions $R(\lambda)$ and $S(y,\lambda)$ are defined as
\begin{equation}
R(\lambda)=\displaystyle\prod_{j=0}^{2g}(\lambda-\lambda_j),\hspace{4mm}
S(y,\lambda)=\displaystyle\prod_{k=1}^{g}(\lambda-{\gamma}_k(y)).
\end{equation}
The multi-valued function $p(P)=-\frac{i}{\Pi}\ln \mu(P)$ is called the {\it
quasi-momentum}. Its differential
$$
d p=-\frac{i}{\Pi}\frac{d\mu}{\mu}
$$
is the unique meromorphic $1$-form on $\Gamma$ satisfying the following properties:

\begin{enumerate}
\item $d p$ has a single pole of second order at
infinity with the principal part
$$
d p= \left(-\frac{1}{k^2}+O(1)\right)d k,
$$
where $k=\lambda^{-1/2}$ is the local parameter, and

\item The periods of $d p$ over the $a$-cycles are equal to zero:
\begin{equation}
\displaystyle\oint_{a_k}d p=0, \hspace{4mm}k=1,\ldots,g
\end{equation}

\end{enumerate}

Since the function $\mu(\lambda)=e^{i \Pi p(\lambda)}$ is
single-valued, the $b$-periods of $\Omega_p$ are integral multiples of
$2 \pi/\Pi$:
\begin{equation}
\displaystyle\oint_{b_k}d p= \frac{2\pi n_k}{\Pi},\hspace{4mm}n_k\in \mathbb{Z},\hspace{4mm}k=1,\ldots,g.
\end{equation}

Conversely, given a hyperelliptic Riemann surface $\Gamma$ together with a
two-sheeted covering $\lambda:\Gamma\rightarrow\mathbb{C P}^1$
with real-valued ramification points $\lambda_0,\ldots,\lambda_{2g}$ and $\infty$, and
a nonspecial divisor $D=P_1+\cdots+P_g$ satisfying
$\lambda(P_i)\in[\lambda_{2j-1},\lambda_{2j}]$, there exists a smooth
real-valued potential $u(y)$ of the Schr\"odinger operator with spectral
data $\{\lambda_i,P_j(0)\}$. This potential is given in terms of the
theta-function of $\Gamma$ by the Matveev-Its formula:
\begin{equation}
u(y)=-2\partial^2_y
\ln\theta(y\vec{U}_1-\vec{A}(P_1)-\cdots-\vec{A}(P_g)-\vec{K}|B_{ij})+C(\Gamma),
\label{MatveevIts}
\end{equation}
where $\vec{A}$ is the Abel map, $\vec{K}$ is the vector of
Riemann constants, $C(\Gamma)$ is a constant, and the vector $\vec{U}_1$ is the vector of $b$-periods
of the unique meromorphic differential $d p$ on $\Gamma$ satisfying properties (1)
and (2) above:
\begin{equation}
(\vec{U}_1)_k=\frac{1}{2\pi}\displaystyle\oint_{b_k}d p\hspace{4mm} k=1,\ldots,g.
\end{equation}
This potential is periodic with period $\Pi$
only if the components of the vector $\vec{U}_1$ are integral multiples of $1/\Pi$.
However, for generic spectral data the components of $\vec{U}_1$ are
arbitrary real numbers, so the potential $u(y)$ is in general
quasi-periodic.

Therefore, the problem of describing all finite-gap potentials of period
$\Pi$ is reduced to the following: describe all
hyperelliptic Riemann surfaces such that the $1$-form $d p$,
uniquely determined by conditions (1) and (2), has $b$-periods
that are integral multiples of $2\pi/\Pi$. An effective solution of this
problem was given in \cite{Grinevich}, which we now recall.

\section{Isoperiodic deformations for the Schr\"odinger operator}

\setcounter{equation}{0}

Let $\Gamma_0$ be a hyperelliptic Riemann surface corresponding to a
real-valued potential of the Schr\"odinger operator of period one.
Consider a deformation $\Gamma(t)$ of $\Gamma_0$, i.e.~let a
continuously varying one-periodic family of hyperelliptic Riemann surfaces
of genus $g$, equipped with two-sheeted covering maps $\lambda(t):\Gamma(t)\rightarrow
\mathbb{C P}^1$, branched at points $\lambda_0(t),\ldots,\lambda_{2g}(t)$
on the real axis and $\infty$, such that $\Gamma(0)=\Gamma_0$. If each
of the curves $\Gamma(t)$ also corresponds to a potential of period one,
then this deformation is called {\it isoperiodic}. The principal result of
\cite{Grinevich} consists of an effective description of all such
deformations.

Suppose we have an isoperiodic deformation $\Gamma(t)$, then each of the curves
has a unique meromorphic $1$-form $d p(t)$ satisfying properties (1) and
(2) above, such that its $b$-periods are integers. Consider the meromorphic $1$-form
\begin{equation}
\omega=\frac{\partial p}{\partial t}d \lambda-\frac{\partial
\lambda}{\partial t}d p
\end{equation}
on $\Gamma_0$. This form has a double pole at infinity and no other
singularities, such forms are called weakly meromorphic. If we choose
the connection in such a way that
$\frac{\partial p}{\partial t}=0$, then the deformation is explicitly
given in terms of the ramification points
\begin{equation}
\frac{\partial \lambda_j}{\partial t}=-\frac{\omega(\lambda_k)}{d
p(\lambda_k)}.
\end{equation}
Conversely, given a weakly meromorphic $1$-form $\omega$,
we can define a deformation of $\Gamma_0$ using the above
formula. If we now choose the connection in such a way that
$\frac{\partial \lambda}{\partial t}=0$, then we see that
\begin{equation}
\frac{\partial p}{\partial t}=\frac{\omega}{d \lambda}
\end{equation}
is a single-valued function on $\Gamma_0$, so therefore the periods
of $d p(t)$ are constant. Therefore, if $\Gamma_0$ corresponds to a
potential of period one, then so do the surfaces $\Gamma(t)$. Thus,
isoperiodic deformations of the Schr\"odinger operator are described
by meromorphic $1$-forms with prescribed singularities, namely with a
double pole at infinity. We now try to adapt this approach to obtain
isoperiodic deformations for the acoustic operator.

\section{Spectral theory of the acoustic operator}

\setcounter{equation}{0}

We now consider the acoustic operator
\begin{equation}
L_A=-r^2(x)\frac{d^2}{d x^2}, \label{AcOp}
\end{equation}
where the function $r(x)$, called the density, is smooth and positive. The
spectral theory of the acoustic operator with a smooth periodic density has been
widely studied. The $L^2(\mathbb{R})$ spectrum of the problem
\begin{equation}
L_A\psi(x,\lambda)=\lambda\psi(x,\lambda),
\end{equation}
like
that of the periodic Schr\"odinger operator, has a zone structure and
consists of an infinite number of bands
$[\Lambda_0,\Lambda_1],[\Lambda_2,\Lambda_3],\ldots$, with
$\Lambda_{2j}<\Lambda_{2j+1}\leq\Lambda_{2j+2}$. If the density $r(x)$ is
smooth, then there is a constraint $\Lambda_0=0$.
 The finite-gap densities
of $L_A$ can be constructed using the following well-known relation
\cite{Nucci}:
\begin{proposition}
Suppose $r(x), \psi(x,\lambda)$ satisfy the acoustic equation
\begin{equation}
-r^2(x)\frac{d^2}{d x^2}\psi(x,\lambda)=\lambda\psi(x,\lambda).
\end{equation}
Perform a change of variables
\begin{equation}
y(x)=\displaystyle\int_0^x\frac{d x'}{r(x')}. \label{Hopf1}
\end{equation}
Then the functions
\begin{equation}
u(y)=\frac{1}{4}r^2_x(x)-\frac{1}{2}r(x)r_{xx}(x),
\label{Hopf2}
\end{equation}
\begin{equation}
\varphi(y,\lambda)=\psi(x,\lambda)r^{-1/2}(x),
\label{Hopf3}
\end{equation}
satisfy the Schr\"odinger equation
\begin{equation}
-\frac{d^2}{d
x^2}\varphi(y,\lambda)+u(y)\varphi(y,\lambda)=\lambda\varphi(y,\lambda).
\end{equation}
If $r(x)$ is a periodic density of $L_A$ with period $T$, then $u(y)$ is a periodic
potential of $L_S$ with period
\begin{equation}
\Pi=\displaystyle\int_{x_0}^{x_0+T}\frac{d x}{r'(x)},
\end{equation}
and the operators $L_A$ and $L_S$ have the same
spectrum in $L^2(\mathbb{R})$.
\end{proposition}

This proposition allows us to construct finite-gap densities of the
acoustic operator by taking a finite-gap potential $u(y)$ of the Schr\"odinger
operator, given by the Matveev-Its formula for some spectral data
$\{\lambda_i,\gamma_k\}$ with $\lambda_0=0$,
and solving the equations
(\ref{Hopf1})-(\ref{Hopf2}) for the density $r(x)$.
This inverse transformation is only defined up to a parameter, as the acoustic
equation has the gauge transformation
\begin{equation}
x\rightarrow\alpha x,\hspace{4mm}
r(x)\rightarrow \alpha^{-1}r(\alpha x),
\label{Gauge}
\end{equation}
so that a periodic potential corresponds to a one-parameter family
of periodic densities. The choice of this $\alpha$ should be considered as an
additional spectral parameter of the problem.

Explicit formulas for $r(x)$ in
terms of theta-functions were obtained by Dmitrieva in
\cite{Dmitrieva1}-\cite{Dmitrieva3} by extending the Hopf transformation
(\ref{Hopf1})-(\ref{Hopf3}) to a transformation between the Harry Dym
hierarchy and the Korteweg-de Vries hierarchy. Finite-gap periodic
densities of the acoustic operator are then equivalent to $x$-periodic
solutions of the Harry Dym equation
\begin{equation}
r_t=r^3r_{xxx}.
\end{equation}

Since periodic densities correspond to periodic potentials, and
periodic potentials of the Schr\"odinger operator are described by the isoperiodic deformations,
the problem of describing all periodic densities of the acoustic
operator is in some sense solved. However, we would like to find
a natural choice of the constant for the gauge transformation, depending
explicitly on the spectral data, and
a set of differential equations on the spectral data of
the acoustic operator, such that the period of the unique density
corresponding to the spectral data is preserved under these flows. We
now turn to this problem.

Let $r(x)$ be a finite-gap density of period $T$,
and let $\psi_{\pm}(x,x_0,\lambda)$ be the
Bloch-Floquet function of the corresponding acoustic operator $L_A$, that is a
joint eigenfunction of the acoustic and monodromy operators:
\begin{equation}
L_A\psi_{\pm}(x,x_0,\lambda)=\lambda\psi_{\pm}(x,x_0,P),
\end{equation}
\begin{equation}
\psi_{\pm}(x+T,x_0,\lambda)=e^{\pm iTq(\lambda)}\psi_{\pm}(x,x_0,\lambda),
\end{equation}
normalized by the relation
\begin{equation}
\left.\psi_{\pm}(x,x_0,T)\right|_{x=x_0}=1.
\end{equation}
Let $L_S$ be the associated Schr\"odinger operator given by
(\ref{Hopf1})-(\ref{Hopf3}). Then the function $\psi(x,x_0,\lambda)$ can
be expressed in terms of the Bloch-Floquet function
(\ref{BlochFloquet1})-(\ref{BlochFloquet2}) of $L_S$ as follows:
\begin{equation}
\psi_{\pm}(x,x_0,\lambda)=r^{-1/2}(x_0)r^{1/2}(x)\varphi(y,y_0,\lambda(P)).
\end{equation}
Hence, we can consider the Bloch-Floquet as a meromorphic function on the
spectral curve $\Gamma$ of $L_S$, i.e.~we can set
\begin{equation}
\psi_{\pm}(x,x_0,\lambda)=\psi(x,x_0,\lambda(P))
\end{equation}
for some meromorphic function $\psi(x,x_0,P)$ on $\Gamma$ satisfying
\begin{equation}
L_A\psi(x,x_0,P)=\lambda(P)\psi(x,x_0,P),
\end{equation}
\begin{equation}
\psi(x+T,x_0,P)=e^{iTq(P)}\psi(x,x_0,P),
\end{equation}
where the multivalued function $q(P)$ is called the {\it quasi-momentum}
of the acoustic operator $L_A$. It can be expressed in terms of the logarithmic derivative of the
Bloch-Floquet function
\begin{equation}
\xi(x,P)=-i\frac{\psi_x(x,x_0,P)}{\psi(x,x_0,P)}
\end{equation}
as follows:
\begin{equation}
q(P)=\frac{1}{T}\displaystyle\int_{x_0}^{x_0+T}\xi(x,P)d x.
\end{equation}
To study the function $\xi(x,P)$, we first see that it can be expressed in
terms of the logarithmic derivative $\chi(y,P)$ of $\varphi(y,y_0,P)$ as
follows:
\begin{equation}
\xi(x,P)=\frac{1}{r(x)}\chi(y,P)-\frac{i}{2}\frac{r_x(x)}{r(x)}.
\end{equation}
Substituting this in (\ref{LogBF}), we get
\begin{equation}
\xi(x,P)=\frac{\sqrt{R(\lambda)}}{r(x)S(x,\lambda)}-\frac{i}{2}
\frac{r'(x)}{r(x)}-\frac{i}{2}\frac{S_x(x,\lambda)}{S(x,\lambda)}, \label{XiExpr}
\end{equation}
where $R(\lambda)=\lambda\displaystyle\prod_{j=1}^{2g}(\lambda-\lambda_j)$
and $S(x,\lambda)=\displaystyle\prod_{k=1}^{g}(\lambda-\gamma_k(y(x)))$.
Therefore, the quasi-momentum $q(P)$ and its differential
have the following high-energy expressions:
\begin{equation}
q(P)=\frac{1}{kT}\displaystyle\int_{x_0}^{x_0+T}\frac{d x}{r(x)}+O(1),
\label{Qatinfty}
\end{equation}
\begin{equation}
d q(P)=\left(-\frac{1}{k^2T}\displaystyle\int_{x_0}^{x_0+T}\frac{d
x}{r(x)}+O(1)\right)d k,
\label{DQatinfty}
\end{equation}
where $k=\lambda^{-1/2}$.

A simple calculation also shows that $\xi(x,P)$ satisfies the Ricatti equation:
\begin{equation}
-i\xi'(x,P)+\xi^2(x,P)-\frac{\lambda(P)}{r^2(x)}=0.
\label{AcRiccati}
\end{equation}
The spectrum of the acoustic operator always starts at $\lambda=0$,
i.e.~the leftmost branch point of $\Gamma$ is $\lambda=0$.
We consider the Ricatti equation
in the neighborhood of this point. Let
$k=\sqrt{\lambda}$ be the local parameter
and consider the Taylor series for $\xi(x,P)$ at $k=0$:
\begin{equation}
\xi(x,P)=\xi_0(x)+\xi_1(x)k+\xi_2(x)k^2+O(k^3).
\end{equation}
Substituting this into the Ricatti equation, we get that
$\xi_0(x)=0$, $\xi_1(x)=C$ for some constant $C$, and that
$\xi_2(x)$ is a full derivative. Therefore, the quasi-momentum $q(P)$
and its differential near $k=0$
are equal to
\begin{equation}
q(P)=C k+O(k^3).
\end{equation}
\begin{equation}
d q(P)=(C+O(k^2))d k.
\end{equation}
On the other hand, comparing the obtained expression for $\xi(x,P)$ with (\ref{XiExpr}), we
see that
\begin{equation}
C=\frac{\sqrt{\lambda_1\cdots\lambda_{2g}}}{(-1)^g r(x)
\gamma_1(x)\cdots\gamma_g(x)},
\end{equation}
and hence $r(x)$ is expressed in terms of the spectral data and the
additional constant $C$ as follows:
\begin{equation}
r(x)=\frac{\sqrt{\lambda_1\cdots\lambda_{2g}}}{(-1)^g C
\gamma_1(x)\cdots\gamma_g(x)}.
\label{Rexpr}
\end{equation}
The constant $C$ in this formula corresponds to the gauge transformation
(\ref{Gauge}) and should be considered in addition to the $\lambda_j$ and
$\gamma_k$ as part of the spectral data of the problem.

To construct isoperiodic deformations of the acoustic operator, we
need a rule of choosing the constant $C$. The natural choice seems to be
$C=(-1)^g$. With this choice, the value of $d q$ at $\lambda=0$ does not
depend on the spectral curve $\Gamma$, which will be used to construct the
isoperiodic deformations. Also, with this choice of $C$, a vanishingly
small potential $u(x)\rightarrow 0$, or equivalently vanishingly small
energy gaps
$(\lambda_{2j-1}-\lambda_{2j}\rightarrow 0)$, correspond to a density
$r(x)\rightarrow 1$ normalized at unity.

We summarize the results of this section in the following
\begin{theorem}
Let $\left\{\lambda_i,\gamma_k(y)\right\}$ be the spectral data of a
finite-gap periodic Schr\"odinger operator $L_S$ with potential $u(y)$
such that $\lambda_0=0$. Then the
density $r(x)$ of the associated acoustic operator $L_A$ related to $u(y)$ by
(\ref{Hopf1})-(\ref{Hopf3}) is given by the equation
(\ref{Rexpr}), where $C$ is an arbitrary constant. If we choose
$C=(-1)^g$, then the differential of the quasi-momentum of $L_A$ has a
pole of second order at $\lambda=\infty$ with principal part
(\ref{DQatinfty}), and has fixed first and second order terms at
$\lambda=0$:
\begin{equation}
d q(P)=((-1)^g+O(k^2))d k.
\end{equation}
\end{theorem}

Using this theorem, we construct isoperiodic deformations for the
acoustic operator.

\section{Isoperiodic deformations of the acoustic operator}

\setcounter{equation}{0}

We recall that a deformation of a Riemann surface is given by
a meromorphic $1$-form
\begin{equation}
\omega=\frac{\partial q}{\partial t}d \lambda.
\end{equation}
To give an isoperiodic deformation for the acoustic operator, this
$1$-form must satisfy the following conditions:

\begin{enumerate}

\item The quasimomentum differential $d q$ has a second order pole
at infinity, so $\frac{\partial q}{\partial t}$ has a first order pole at infinity.
Since $d\lambda$ has a third order pole at infinity, $\omega$ has
a pole of fourth order at infinity.

\item At zero $d q$ has fixed first and
second order terms, therefore $\frac{\partial q}{\partial t}$ has a third order zero
(with our choice of $C$). Since $d\lambda$ has a first order zero,
$\omega$ has a zero of fourth order at $\lambda=0$.

\end{enumerate}

Therefore, a $1$-form $\omega$ defines an isoperiodic deformation
if and only if it has divisor $4\cdot 0-4\cdot\infty$. The space of such forms
is $g$-dimensional. Since the total space of spectral curves has dimension
$2g$, and there are $g$ conditions for the period to be equal to one,
these forms are a basis of isoperiodic deformations.

As before, the deformation is given explicitly in terms of the branch points as
\begin{equation}
\frac{\partial \lambda_j}{\partial t}=-\frac{\omega(\lambda_j)}{d
q(\lambda_j)},\hspace{4mm}j=1,\ldots,2g.
\end{equation}

On a hyperelliptic surface, the $1$-form $d q$ can be written
explicitly as
\begin{equation}
d q=\frac{Q(\lambda)}{2\sqrt{R(\lambda)}}d\lambda,
\end{equation}
where
\begin{equation}
Q(\lambda)=q_{g}\lambda^{g}+\cdots+q_0,
\end{equation}
and $q_0=(-1)^g\sqrt{\lambda_1\cdots\lambda_{2g}}$. The
coefficients of the polynomial $Q(\lambda)$ are determined by
setting the $a$-periods of $d q$ to zero and are expressed in
terms of certain hyperelliptic integrals.

An arbitrary $1$-form $\omega$ with divisor $4\cdot
0-4\cdot\infty$ can be written as
\begin{equation}
\omega=\frac{f(\lambda)}{2\sqrt{R(\lambda)}}d\lambda,
\end{equation}
where
\begin{equation}
f(\lambda)=f_{g+1}\lambda^{g+1}+\cdots+f_2\lambda^2=\lambda^2(f_{g+1}\lambda^{g-1}+\cdots+f_2).
\end{equation}

The deformation given by such an $\omega$ has the form
\begin{equation}
\frac{\partial \lambda_j}{\partial
t}=-\frac{f(\lambda_j)}{Q(\lambda_j)},\hspace{4mm}j=1,\ldots,2g.
\end{equation}

It is possible to choose a basis in the space of meromorphic $1$-forms
with divisor $4\cdot 0-4\cdot\infty$ and construct a basis of
deformations by this formula. However, the right-hand side of the
deformation equations contains the coefficients of $Q(\lambda)$
that are in turn expressed as hyperelliptic integrals containing
the ramification points. To avoid this difficulty, we use the approach of
\cite{Grinevich}. Factoring the polynomial $Q(\lambda)$
\begin{equation}
Q(\lambda)=(\beta_1\lambda-\sqrt{\lambda_1\lambda_2})\cdots
(\beta_g\lambda-\sqrt{\lambda_{2g-1}\lambda_{2g}}),
\end{equation}
we consider the parameters $\beta_k$ as independent
variables that are deformed along with the ramification points.
This extension of the phase space greatly simplifies the
deformation equations.

The deformations of $\beta_k$ are easily shown to be given by the
following
\begin{lemma}
Let $\frac{\partial}{\partial t}$ be the flow given by the form $\omega$
with polynomial $f(\lambda)$. Then the deformations of $\lambda_i$ and
$\beta_k$ have the form
\begin{equation}
\frac{\partial \lambda_j}{\partial t}=-\frac{f(\lambda_j)}{Q(\lambda_j)}
\end{equation}
\begin{eqnarray}
\frac{\partial \beta_k}{\partial
t}=\beta_k\left[-\frac{1}{2\lambda_{2k-1}}\frac{f(\lambda_{2k-1})}{Q(\lambda_{2k-1})}
-\frac{1}{2\lambda_{2k}}\frac{f(\lambda_{2k})}{Q(\lambda_{2k})}
+\right.\nonumber\\
+\frac{1}{Q'(\lambda)}
\cdot\left.\left.\left(f'(\lambda)-\frac{f(\lambda)R'(\lambda)}{2R(\lambda)}\right)
\right|_{\lambda=\frac{\sqrt{\lambda_{2k-1}\lambda_{2k}}}{\beta_k}}\right]
\end{eqnarray}
\end{lemma}

As we see, the equations no longer contain hyperelliptic
integrals. This form makes the equations of isoperiodic
deformation useful for numerical simulations.

To write the deformations explicitly, we choose a basis in the
space of differentials with divisor $4\cdot 0-4\cdot\infty$ as
follows:
\begin{equation}
\omega_k=\frac{c_k\lambda^2}{\beta_k\lambda-\sqrt{\lambda_{2k-1}\lambda_{2k}}}d
q,
\label{Omegas}
\end{equation}
where $c_k$ are arbitrary constants. Then
\begin{equation}
\frac{f_k(\lambda)}{Q(\lambda)}=\frac{c_k\lambda^2}{\beta_k\lambda
-\sqrt{\lambda_{2k-1}\lambda_{2k}}},
\end{equation}
and the deformation equations are given by our final theorem.

\begin{theorem}
Let $\Gamma$ be a hyperelliptic Riemann surface with ramification points
$\lambda_0=0,\lambda_1,\ldots,\lambda_{2g}$, corresponding to a smooth
density of period one. Let the zeroes of the quasimomentum be
$\frac{\sqrt{\lambda_{2k-1}\lambda_{2k}}}{\beta_k}, k=1,\ldots,g$.
Consider the flows
\begin{equation}
\frac{\partial \lambda_j}{\partial
t_k}=-\frac{c_k\lambda^2_j}
{\beta_k\lambda_j-\sqrt{\lambda_{2k-1}\lambda_{2k}}}
\end{equation}
\begin{eqnarray}
\frac{\partial \beta_l}{\partial
t_k}=c_k\beta_k\left[-\frac{1}{2\lambda_{2l-1}}
\frac{\lambda_{2l-1}^2}{\beta_k\lambda_{2l-1}-\sqrt{\lambda_{2k-1}\lambda_{2k}}}-
\frac{1}{2\lambda_{2l}}
\frac{\lambda_{2l}^2}{\beta_k\lambda_{2l}-\sqrt{\lambda_{2k-1}\lambda_{2k}}}+\right.
\nonumber\\
\left.+\frac{\lambda_{2l-1}\lambda_{2l}}{\beta_l^2}\left(
\beta_k\frac{\sqrt{\lambda_{2l-1}\lambda_{2l}}}{\beta_l}-
\sqrt{\lambda_{2k-1}\lambda_{2k}}\right)^{-1}\right];\hspace{4mm}l\neq k\hspace{15mm}
\end{eqnarray}
\begin{eqnarray}
\frac{\partial \beta_k}{\partial
t_k}=c_k\beta_k\left[-\frac{1}{2\lambda_{2k-1}}
\frac{\lambda_{2k-1}^2}{\beta_k\lambda_{2k-1}-\sqrt{\lambda_{2k-1}\lambda_{2k}}}-
\frac{1}{2\lambda_{2l}}
\frac{\lambda_{2k}^2}{\beta_k\lambda_{2k}-\sqrt{\lambda_{2k-1}\lambda_{2k}}}+\right.\nonumber\\
+\frac{1}{\beta_k}\left\{\frac{\lambda_{2k-1}\lambda_{2k}}{\beta_k^2}
\displaystyle\sum_{l\neq k}\beta_l\left(
\beta_l\frac{\sqrt{\lambda_{2k-1}\lambda_{2k}}}{\beta_k}-
\sqrt{\lambda_{2l-1}\lambda_{2l}}\right)^{-1}+
2\frac{\sqrt{\lambda_{2k-1}\lambda_{2k}}}{\beta_k}-\right.\nonumber\\
\left.\left.
\frac{\lambda_{2k-1}\lambda_{2k}}{\beta_k}\cdot
\left(\frac{\beta_k}
{\sqrt{\lambda_{2k-1}\lambda_{2k}}}-\displaystyle\sum_{j=1}^{2g}
\frac{\beta_k}{\sqrt{\lambda_{2k-1}\lambda_{2k}}-\lambda_j\beta_k}\right)\right\}\right]\hspace{15mm}
\end{eqnarray}
on the spectral data,
where $\frac{\partial}{\partial t_k}$ is the flow given by $\omega_k$
defined by (\ref{Omegas}). Then any Riemann surface obtained by moving
along these flows in the moduli space of hyperelliptic Riemann surfaces
corresponds to a density $r(x)$ of period one of the acoustic operator.
These flows form a basis in the
space of isoperiodic deformations of the acoustic operator.

\end{theorem}

We see that these extended equations do not involve
hyperelliptic integrals.

\section{Acknowledgements}

This work was done in part at the Los Alamos Summer Program in
Mathematical Modeling and Analysis in 2004. The author would like to
sincerely thank Prof. P. G. Grinevich for
supervising the author's work, and Mikhail Stepanov for many instructive
discussions.

\end{document}